# Room temperature ferromagnetism and anomalous Hall effect in $Si_{1-x}Mn_x$ ($x \approx 0.35$) alloys

B.A. Aronzon[1,8,a)], V.V. Rylkov[1,9,b)], S.N. Nikolaev[1], V.V. Tugushev[1,c)], S. Caprara[2], V.V. Podolskii[3], V.P. Lesnikov[3], A. Lashkul[4], R. Laiho[5], R.R. Gareev[6], N.S. Perov[7], A.S. Semisalova[7]

[1]Russian Research Centre "Kurchatov Institute", Moscow, 123182, Russia

[2]Dipartimento di Fisica, Universita di Roma "La Sapienza", piazzale Aldo Moro, 2 - 00185 Roma, Italy

[3]Physicotechnical Research Institute, Nizhnii Novgorod State University, Nizhnii Novgorod, 603950, Russia

[4]Lappeenranta University of Technology, Lappeenranta, P.O. box 20, 53850, Finland

[5]Wihuri Physical Lab., Turku University, Turku, 20014, Finland

[6]Institute of Experimental and Applied Physics, University of Regensburg, 93040 Regensburg, Germany

[7]Faculty of physics, M.V. Lomonosov MSU, Moscow, 119992, Russia

[8]Institute of Applied and Theoretical Electrodynamics, Russian Academy of Sciences, Moscow, 127412, Russia

[9]Kotel'nikov Institute of Radio Engineering and Electronics, Russian Academy of Sciences, Fryazino, Moscow District 141190, Russia

---

a)Electronic mail: aronzon@imp.kiae.ru

b)Electronic mail: vvrylkov@mail.ru

c)Electronic mail: tuvictor@mail.ru

# ABSTRACT

A detailed study of the magnetic and transport properties of $Si_{1-x}Mn_x$ ($x \approx 0.35$) films is presented. We observe the anomalous Hall effect (AHE) in these films up to room temperature. The results of the magnetic measurements and the AHE data are consistent and demonstrate the existence of long-range ferromagnetic (FM) order in the systems under study. A correlation of the AHE and the magnetic properties of $Si_{1-x}Mn_x$ ($x \approx 0.35$) films with their conductivity and substrate type is shown. A theoretical model based on the idea of a two-phase magnetic material, in which molecular clusters with localized magnetic moments are embedded in the matrix of a weak itinerant ferromagnet, is discussed. The long-range ferromagnetic order at high temperatures is mainly due to the Stoner enhancement of the exchange coupling between clusters through thermal spin fluctuations ("paramagnons") in the matrix. Theoretical predictions and experimental data are in good qualitative agreement.

# 1. Introduction

Materials which display both high-temperature ferromagnetism and good semiconducting properties are of great interest for basic research and respond to the challenging search for electronics applications in which both the charge and the spin of the electrons are used for information storage and processing [1]. Materials which consist of magnetic transition metals embedded in nonmagnetic semiconductor, as a matrix, are the most promising to this purpose. It is widely accepted that such materials could be used for injection of spin-polarized carriers into a normal semiconductor at temperatures above room temperature [1]. Up to now the related studies were mainly oriented to dilute magnetic semiconductors (DMSs) based on Mn-doped III-V semiconducting materials (III-Mn-V, mostly GaMnAs). In such materials, if the Mn concentration is not too high, Mn substitutes Ga acting as an acceptor, so doping GaAs with Mn yields both local magnetic moments and free holes [2, 3]. The ferromagnetic (FM) ordering in this case is due to an indirect exchange between Mn atoms accompanied by the spin polarization of holes which could reach 80% [4, 5]. On the other hand, there are quite few papers dealing with studies FM ordering in Mn doped Si in spite of the fact that such structures are mostly interesting, due to the compatibility with mainstream silicon technology.

It is known [6] that isolated Mn impurities occupy mainly tetrahedral interstitial positions ($Mn_T^-$, $Mn_T^0$, $Mn_T^+$, $Mn_T^{2+}$) in the Si lattice acting as donors [2, 3], while a strong hybridization of Mn 3d-states with 4(s,p)-states in Si and an indirect exchange between Mn moments appears if they enter into substitution positions ($Mn_{Si}^{2-}$, $Mn_{Si}^+$) as acceptors [2, 7]. The combined simultaneous deposition of Mn and Si could lead to the formation of manganese silicides $Mn_nSi_m$ films, with the ratio ($m/n$) ranging between 1.70 and 1.75 (for example, $Mn_4Si_7$, $Mn_{11}Si_{19}$, $Mn_{15}Si_{26}$, $Mn_{26}Si_{45}$, $Mn_{27}Si_{47}$) [8]. These phases may exhibit semiconducting, metallic, or half-metallic behavior [9]. For example, ideally stoichiometric and unstressed $Mn_4Si_7$ is shown to be a semiconductor with an indirect band gap, although a small non-stoichiometry or lattice stress lead to the closure of the gap, turning the semiconductor into a semimetal or a metal. It is important that some manganese silicides [8-10] are weak itinerant ferromagnets with Curie temperatures $T_c < 50$ K. It should be also noted that electron transport in $Mn_nSi_m$ films is not well known, and to our knowledge resistivity and Hall effect in $Mn_4Si_7$ are studied only in Ref. [8].

Si based DMSs attracted interest after the observation of a FM state with high Curie temperature, $T_c > 400$ K, in these materials [11]. This result was obtained in Si:Mn films with a relatively low (0.1 – 0.8 at.%) content of implanted Mn ions, but its origin remained mysterious. Since a FM state was also observed in Si after implantation of nonmagnetic ions (Ar, Si) or irradiated by neutrons, some authors argued that high-temperature ferromagnetism in Si based DMSs is due to paramagnetic defects [12, 13].

Detailed X-ray and magnetic studies of Mn implanted Si indicated that the Mn ions enter not only in the substitution or interstitial positions of the Si lattice, but also form molecular clusters [13-15]. Another assumption [16] was that the FM signal is due to such clusters arising during the growth processes. Thus, the Si based DMSs seem to be very inhomogeneous alloys and their magnetic properties strongly differ from those of bulk $Mn_nSi_m$ materials. Furthermore, formation of the bulk $Mn_nSi_m$ precipitate particles in these DMSs could not be itself the reason for FM ordering, because, as we remarked above, the $Mn_nSi_m$ silicides have their Curie temperature $T_c$ < 50 K. It could also be mentioned that the effective magnetic moment per Mn atom (> 0.2 $\mu_B$/Mn) in Mn-implanted Si [15, 18] is quite large as compared with the one in the bulk $Mn_nSi_m$ (for example, ≈0.012 $\mu_B$/Mn in $Mn_4Si_7$ [8]).

Recently, it was calculated [17] that stable FM Ångstrom sized Mn-Si complexes can appear in the Si matrix; they contain 2 or 3 Mn atoms (dimers or trimers) and have effective magnetic moments (2-3) $\mu_B$/Mn. In $Si_{1-x}Mn_x$ alloys, such complexes can also be self-organized in isolated nanometer sized (≈20 nm) $Si_{1-x}Mn_x$ ($x \approx 0.35$) precipitate grains containing some hundreds or even thousands of Mn atoms [15]. So, in the material with a relatively low Mn content, the FM signal is probably not due to the formation of a global FM state of isolated Mn moments coupled via spin–polarized carriers, but could be rather attributed to isolated FM Ångstrom sized complexes or nanometer sized precipitate grains. As a result, Si:Mn systems with a low Mn content attract less interest for spintronics and the main trends concern materials with a relatively high Mn content. However, for these materials the situation is also controversial. For example, a FM state with $T_c \approx$ 250 K was observed in uniformly doped $Si_{1-x}Mn_x$ ($x$ = 0.03-0.05) films prepared by magnetron sputtering followed with fast annealing [19]. High $T_c$ values (~ 300 K) were also recently observed in digital heterostructures with alternating deposition of Mn and Si thin layers with average Mn content 5-10 at.% [20]. On the other hand, amorphous $Si_{1-x}Mn_x$ films with $x$ = 0.005 – 0.175, obtained by a similar method, showed extremely small magnetic moment per Mn atom and the Curie temperature did not exceed 2 K [21].

Thus, we can conclude that the mechanism for FM ordering at $T$ > 50 K in Si doped with Mn is far from understood. Furthermore, the results obtained for samples prepared by similar techniques could contradict one another (compare, for example, the results published in Refs. [11] and [15] or in Refs. [20] and [21]).

The common feature of the main experimental results [11 – 13, 15, 16, 18 - 21] is that they were based on magnetization measurements which could not be the proof of global FM order and spin-polarization of the carriers. For example, the hysteresis loop of magnetization could be observed even at the room temperatures in III-Mn-V DMSs with embedded FM nanograins (MnAs or MnSb), while the anomalous contribution to the Hall effect related to the carrier spin polarization is absent (see Ref. [22] and references therein). To get at least a hint of the carrier spin polarization and to detect the interplay of the magnetic and electronic subsystems, one needs to measure the anomalous Hall effect (AHE) which is proportional to the magnetization and is due to spin–orbit interaction and spin-polarization of carriers in DMSs [2, 3]. Such measurements play a key role in III-Mn-V DMSs for the identification of the FM state [2, 3]. However, in $Si_{1-x}Mn_x$ systems the observation of the AHE displaying a hysteresis loop at sufficiently high temperature (230 K) has been reported up to now only in our paper [23]. Recently, the AHE was also observed in hydrogenated amorphous Si:H with Mn content up $x \approx$ 0.35 at $T \leq 150$ K, while the hysteresis loop of the AHE was absent [24].

In this paper we present the results of magnetic, AHE and resistivity measurements in $Si_{1-x}Mn_x$ films with a high Mn content ($x \approx 0.35$), which demonstrate the global FM order and indicate the carrier spin – polarization at temperatures up to room temperature. A theoretical model is developed to explain the experimental results. This model takes into account the existence of FM Ångstrom sized Mn-Si complexes with localized magnetic moments, embedded in the $Mn_nSi_m$ host, which is a weak itinerant ferromagnet.

The paper is organized as follows. The sample characteristics and experimental methods are described in Sec. 2. In Sec. 3 we present the data of the temperature dependence of the resistivity and the Hall effect, mainly the AHE, measurements. The results of the magnetic measurements are presented in Sec. 4; they are in a good agreement with the AHE data which demonstrate the existence of long-range FM order in the systems under study up to room temperatures. We show a correlation of the AHE and magnetic properties of the studied films with their conductivity and substrate type. In Sec. 5 we discuss the experimental results and develop a theoretical model. In Sec. 6 we compare the experimental data with theoretical predictions and demonstrate their qualitative agreement. A brief summary and conclusion could be found in Sec.7.

## 2. Samples and methods

$Si_{1-x}Mn_x$ films, mainly of thickness $d$ = 40-80 nm, with about 35 *at.%* of Mn content were deposited from laser plasma on $Al_2O_3$ and GaAs substrates under vacuum conditions ($\approx 10^{-6}$ Torr) using two separated Mn and Si targets [25]. We used 99.9% pure Mn and float zone Si with dopant concentration $\leq 10^{13}$ cm$^{-3}$. The Mn content was controlled by Mn and Si flows and its accuracy was about 10%, estimated by means of electron-probe microanalysis. The substrate temperature $T_g$ mainly stabilized at 300°C, although some samples were grown at different $T_g$ and with different values of *d*, to detect how these parameters affect the sample properties. Special attention was paid to the effect of the substrate on transport and magnetic properties of $Si_{1-x}Mn_x$ films. To that purpose, two types of substrates with different lattice parameters were used, $Al_2O_3$ (*a* = 4.76 Å, *b* = 5.12 Å) and GaAs (*a* = 5.66 Å), with crystallographic orientations ($1\bar{1}02$) and (100), respectively. Some characteristics of the studied samples are listed in Table I.

The sample resistivity at room temperature was in the range $(1.3 - 2.3) \cdot 10^{-4}$ Ω·cm, typical for semimetals or heavily doped semiconductors [26]. From the normal Hall effect resistance we found that carriers are of the hole type and their concentration *p*, estimated from value of the Hall resistance, is about $p \approx 2 \cdot 10^{22}$ cm$^{-3}$. Usually, the carrier concentration in DMSs is noticeably smaller then the Mn content $N_{Mn}$, while in our case *p* is close to the manganese concentration $N_{Mn} \approx 2 \cdot 10^{22}$ cm$^{-3}$ (corresponding to $x \approx 0.35$).

As it was mentioned above, to detect FM long-range ordering it is really essential to compare magnetic and AHE results. Both magnetic and AHE measurements were performed in the range 4.2 – 300 K in magnetic field up to 2.5 T. Hall bar samples of size 2×7 mm$^2$ were prepared for transport measurements, SQUID, longitudinal MOKE hysteresis and vibrating sample magnetometer were used for magnetometry. Magnetic measurements were mainly performed with the magnetic field aligned in the sample plane.

Table 1. $Mn_xSi_{1-x}$ sample parameters.

| Sample number | Growth temperature $T_g$, °C | Substrate | Film thickness *d*, nm | $H_c$, Oe at 80 K | AHE at 300 K exceeding (or not) noise level [1] | AHE sign |
|---|---|---|---|---|---|---|
| №1 | 300 | $Al_2O_3$ | 40 | 2900 | < | - |
| №2 | 300 | $Al_2O_3$ | 57 | 2000 | > | - |
| №3 | 350 | $Al_2O_3$ | 55 | 4200 | ~ | - |
| №4 | 300 | GaAs | 80 | 0 | > | + |
| №5 | 300 | GaAs | 50 | 0 | > | + |
| №6 | 200 | GaAs | 75 | 330 | < | + |
| №7 | 300 | GaAs | 300 | 650 | < | - |

1) Noise level converting in Hall resistance is about $10^{-3}$ Ω at 300 K.

## 3. Resistivity and anomalous Hall effect

The temperature dependencies of the resistance $R_{xx}(T)$ for samples prepared on $Al_2O_3$ and GaAs substrates are presented in Figs. 1a and 1b respectively. The $R_{xx}(T)$ dependencies correspond to the metallic type of conductivity and the resistance variation vs temperature is less then 20% with lowering $T$ from 300 down to 80 K. For samples deposited on the $Al_2O_3$ substrate at $T_g$ = 300 °C the ratio $R_{xx}(290K)/R_{xx}(80K)$ is less than 1.1 (samples 1, 2). On the other hand, the same ratio for samples deposited on the GaAs substrate at the same $T_g$ and of the same thickness is about ≈ 1.2 (samples 4 and 5). This difference in the temperature dependence of the resistance for the two types of samples correlates with the sign of the AHE, which is negative for samples deposited on the $Al_2O_3$ substrate and positive for samples deposited on the GaAs substrate. With increasing film thickness the crystal structure affects the sample properties more weakly and for sample 7, deposited on GaAs substrate with thickness 300 nm, $R_{xx}(290K)/R_{xx}(80K) = 1.12$ is closer to the value obtained for samples with $Al_2O_3$ substrate and the sign of the AHE is negative. When the temperature of measurements is lowered further, special attention should be paid to the abrupt fall of the resistance at temperatures less then 40 K which is much more pronounced for samples deposited on the $Al_2O_3$ substrate (see Fig 1). Note, that in our case the value $R_{xx}(T)$ changes by less than a factor of two in the range 4 – 50K, differing drastically from behavior observed in $Mn_4Si_7$ single crystal, where the abrupt fall at $T < 50$ K is absent and $R_{xx}(T)$ changes by a factor 360 in the range 4 – 300 K [8]. Below, we did not discuss the possible origin of this fall, since the low temperature region $T < 50$ K is out of the scope of our work. Furthermore, as it will be pointed later, the difference of the $R_{xx}(T)$ dependencies for samples prepared on various substrates correlates with the variation of AHE and magnetic properties. Nevertheless, the main result obtained, which is observation of the predominant AHE contribution to the Hall effect at room temperature, is valid for both kinds of samples.

Our main task is to achieve a FM state in Si based structures at high enough temperatures, and in particular the spin-polarization of charge carriers, so the main attention should be paid to the interplay of ferromagnetism and electron transport and, in particular, to the anomalous Hall effect. In magnetic materials the Hall resistance $R_H$ is the sum of the normal and anomalous components:

$$R_H d = \rho_{xy} = R_0 B + R_s M ,$$

here $d$ is the sample thickness, $R_0$ and $R_s$ are constants that characterize the strength of the normal and anomalous Hall resistivities, respectively. The normal Hall effect related to the Lorentz force and is proportional to the magnetic induction $B$, whereas the anomalous Hall effect is proportional to magnetization $M$ and is determined by the spin-orbit interaction and the spin polarization of the carriers. $R_s \propto (R_{xx})^\alpha$, where $\alpha = 1$ for the "skew-scattering" mechanism of AHE and $\alpha = 2$ for "intrinsic" and "side-jump" mechanisms [2].

The curves $\rho_{xy}(B)$ for samples 2 and 4 deposited on $Al_2O_3$ and GaAs substrates and having the same ratio $R_{xx}(290K)/R_{xx}(5K) = 1.3$-$1.4$ are shown in Figs. 2 and 3, respectively, at different temperatures up to room temperature. [Fig. 2 shows the magnetic field dependence of the Hall resistivity $\rho_{xy}(B)$ for sample 2 at temperatures 5–100 K at fields up to 2.5 T (Fig. 2a) and at high temperatures ($\leq 300$ K) in fields $\leq 1$ T (Fig. 2b)]. For both samples the Hall resistance depends nonlinearly on the magnetic field, proving the existence of the anomalous components of the Hall effect, while at high fields the crossover to the linear dependence $\rho_{xy} \propto B$ (normal Hall effect) is observed. Based on the data presented in Figs. 2 and 3 one could argue that in all samples the Hall effect is anomalous at temperatures much higher than the temperature of magnetic ordering in $Mn_4Si_7$ (43 K). Furthermore, the anomalous component of the Hall effect is predominant up to room temperature in both types of structures, being however noticeably stronger for samples deposited on GaAs substrate than for samples with $Al_2O_3$ substrate.

It should be stressed that we have observed the AHE hysteresis loop (see Table 1 and Fig. 2). For sample 2 the dependence $\rho_{xy}(B)$ shows a strong enough coercive field $H_c \approx 2$ kOe at $T \approx 100$ K ($H_c = B_c/\mu_0$) and the hysteresis loop persists up to high enough temperature $\approx 230$ K (see Fig. 2b). In previous studies of $Si_{1-x}Mn_x$ structures the anomalous component of the Hall effect was not observed [16, 18, 19], or the AHE did not show a hysteresis loop [24]. To our knowledge, the hysteresis loop of the AHE at such high temperatures is observed here for the first time in DMSs based on Si.

The behavior of the AHE is different for various substrates, its sign being mainly negative for $Si_{1-x}Mn_x$ /$Al_2O_3$ and positive for $Si_{1-x}Mn_x$ /GaAs. Noticeably, also the coercive force for samples prepared on $Al_2O_3$ substrate is much higher then that in $Si_{1-x}Mn_x$ /GaAs samples. For example in sample 4 the hysteresis is absent (see Fig. 3), while in sample 2 the value of $B_c$ is quite large (see Fig. 2). The absence of coercive force for the AHE in $Si_{1-x}Mn_x$ /GaAs is due to the anisotropy of the magnetic moment, which could be aligned within the sample plane while it is not so for samples prepared on $Al_2O_3$. Such a statement is in agreement with the results of magnetization measurements (see Sec. 4). It is known that the substrate strongly affects the magnetic anisotropy of DMS structures, for example the magnetic moment for ($Ga_{1-x}Mn_xAs/GaAs$) is in the sample plane but for the same samples prepared on $In_{0.16}Ga_{0.84}As$ it is perpendicular to this plane [27]. Results for sample 7 do not contradict the above mentioned tendencies because this sample is very thick (300 nm, about 5 times thicker then other samples) and the effect of the substrate on the sample is weaker.

To shed a light on the mechanism of the AHE one could analyze the parametric dependence $R_s(R_{xx})$. The theory predicts $R_s \propto R_{xx}^a$ with $\alpha =1$ for skew-scattering and $\alpha =2$ for side-jump or intrinsic mechanisms. For samples 1 and 2 grown at $T_g = 300°C$, using the temperature as a parameter we estimated from the $\rho_{xy}(R_{xx})$ dependence a value of $\alpha$ which is in the range between 1.1 and 1.3, in agreement with "skew-scattering", while for sample 3 ($T_g = 350°C$) with higher conductivity $\alpha \approx 2$. The crossover from "skew–scattering" to "intrinsic" AHE mechanism with rising conductivity is natural, because the intrinsic mechanism does not depend on scattering. An analogous tendency was observed in $Ga_{1-x}Mn_xAs$ films where the crossover from "skew-scattering" to "intrinsic" mechanism was observed with increasing conductivity [28]. The difference in the sign of the AHE for $Si_{1-x}Mn_x/Al_2O_3$ and $Si_{1-x}Mn_x$ /GaAs is not related to the sign of charge carriers because in both cases they are holes. This difference is due to the AHE mechanisms because the sign of $R_s$ can be positive or negative depending on the material band structure, on the subtle interplay between the orientations of orbital and spin moments, as well as on the character (repulsive or attractive) of scattering potentials [3]. Particularly, the sign of the AHE is positive for Fe, while it is negative for Ni [26].

## 4. Magnetization

Let us now compare the AHE results with those obtained from magnetization measurements. The magnetic field dependencies of the magnetization $M(H)$ for the $Si_{1-x}Mn_x/Al_2O_3$ structure (sample 2; the area of the sample $S \approx 2x4$ mm$^2$) and for the $Si_{1-x}Mn_x$/GaAs structure (sample 4; $S \approx 2x5$ mm$^2$), measured at various temperatures with field parallel to the sample plane are presented in Figs. 4a and 4b, respectively. It is seen from the data presented in Fig. 4a that the magnetization signal is observed up to 300 K. The saturation magnetization at 80 K is $\approx 12$ emu/cm$^3$. The $M(B)$ dependence shows the hysteresis. The coercivity $H_c$ at 80 K is $\approx 1.2$ kOe. At saturation the magnetic moment per Mn atom for these samples is $\approx 0.07$ $\mu_B$/Mn and $\approx 0.03$ $\mu_B$/Mn for $T = 200$ and 300 K, respectively.

The magnetic moment of sample 4 ($Si_{1-x}Mn_x$/GaAs structure) is well observed at room temperature (see Fig. 4b). However, the hysteresis loops for the samples on GaAs substrate are considerably narrower than those found in the $Si_{1-x}Mn_x$ /Al$_2$O$_3$ structures. Therefore, the measured $M(H)$ curves have been fitted with the Langevin function to obtain the coercivity $H_c$ [the measured dependence $M(H)$ at $T=80$ K is shown in Fig. 4b with curve 1', in turn the curves 1-3 correspond to the fitted dependencies $M(H)$ at 80, 200 and 300 K]. Specifically, the value of coercivity $H_c$ obtained with this fitting procedure for sample 4 at 80 K is about 240 Oe. The saturated value of the magnetic moment per Mn atom for this particular sample is $\approx 0.3 \mu_B$/Mn at $T = 200$ K ($\approx 0.08$ $\mu_B$/Mn for 300 K) and exceeds the value for samples 1 and 2. So, the magnetic moment per Mn atom depends on the substrate and for samples prepared on GaAs substrates it is several times larger than for samples on $Al_2O_3$ substrates. The values obtained are more then one order of magnitude larger then the magnetic moment in $Mn_4Si_7$ ($\approx 0.012 \mu_B$/Mn) [8] and 4 times larger than that in $Si_{1-x}Mn_x$ films with lower Mn content (3.6–5.5 at.%) [19], in which the moment is equal to $(0.03 - 0.05)$ $\mu_B$/Mn at 200 K. It is seen that the coercivity of the samples deposited on GaAs is much smaller than for samples with $Al_2O_3$ substrates, while the opposite holds for the value of $M_s$ (see Fig. 4a and 4b). The same fact is valid for the AHE results: samples with GaAs substrates to all practical extent do not show hysteresis but the AHE at saturation is about a factor of 5 larger than in the $Si_{1-x}Mn_x/Al_2O_3$ structures.

For both samples 2 and 4, the coercivity and saturation moment diminish when the temperature increases, as it is expected.

Some magnetization measurements were performed for sample 2 ($Si_{1-x}Mn_x/Al_2O_3$) with the field perpendicular to the sample plane. The results are comparable with those obtained in the parallel field, the saturated $M_s$ magnetic moments are of the same value in both cases. Also the coercivity found from the AHE (perpendicular field) and magnetization (parallel field) measurements are in agreement with each other, for example, at 80 K $H_c \approx 1.2$ kOe for both cases. Using equation $H_c \cong 2K/M_s$, where $K$ is the anisotropy constant, and the experimental data $H_c \sim 10^3$ Oe and $M_s \sim 10$ emu/cm$^3$ (see Fig.4) one can obtain that the anisotropy is weak, $K \sim 5 \cdot 10^3$ erg/cm$^3$, while the shape anisotropy is even much smaller, being determined by $\sim M_s^2 = 10^2$ erg/cm$^3$. Based on the data presented above, it is natural to suggest that the sample structure consists of crystallites with uniaxial anisotropy which are randomly oriented, resulting in a nearly isotropic behavior of the sample. Growth parameters affect the crystallite anisotropy, for example for the sample 3 $M_r/M_s \approx 1$, while for structures with GaAs substrate (samples 4 – 6) the coercivity is practically absent for the AHE, whereas a small coercivity exists for the magnetic moment (compare Figs. 4a and 4b). The latter could be due to the alignment of the magnetic moment in the sample plane for this particular structure. With increasing sample thickness the influence of the substrate becomes weaker. Accordingly, the sample 7 grown on GaAs substrate with $d \approx 300$ nm displays a quite large $H_c$, determined from the magneto-optical Kerr effect measurements (see below), and negative AHE sign, contrary to other $Si_{1-x}Mn_x$/GaAs samples.

The above mentioned results of magnetization measurements are in agreement with the $\rho_{xy}(B)$ dependence as it may be seen from comparison of the results presented in Fig. 2 with Fig. 4a and Fig. 3 with Fig. 4b. The magnetic field dependences of $\rho_{xy}(B)$ and $M(B)$ are close to each other and show similar hysteresis. In particular, for the sample 2 the coercivity measured by magnetization measurements is approximately of the same as that obtained from AHE. Furthermore, the temperature dependences of the coercivity obtained from transport and magnetic measurements also agree very closely, as it may be seen from Fig. 5a, where the temperature dependences of the normalized coercivity $H_c(T)/H_c(0)$ obtained from both transport and magnetic measurements are presented ($H_c(0)$ is the low temperature value).

The temperature dependence of the saturation magnetization $M_s(T)/M_s(0)$ measured at $B = 1$ T is shown in Fig. 5b. Here, $M_s(0)$ is the $M_s$ value measured at low temperatures. In this figure, the analogous data extracted from the AHE measurements are also shown. To obtain the ratio between the saturation magnetization at temperature $T$ and at zero temperature from the AHE results one should take into account that AHE resistance $R_H^A = R_s M$, where $R_s$ also depends on temperature following the temperature dependence of $R_{x,x}$, i.e., $R_s \propto (R_{xx})^\alpha$, where in our case mainly $\alpha \cong 1$. Hence, we have

$$R_{Hs}^A(T)/R_{Hs}^A(0) = [R_s(T)/R_s(0)] \times [M_H(T)/M_H(0)] = [R_{xx}(T)/R_{xx}(0)]^\alpha \times [M_H(T)/M_H(0)],$$

where the index $H$ for $M$ means that the magnetic moment is extracted from the AHE measurements. The temperature dependence of $M_{sH}(T)/M_{sH}(0)$, where $M_{sH}$ is $M_H$ saturated value, is presented in Fig. 5b and compared with $M_s(T)/M_s(0)$.

The Curie temperature for sample 2 could be estimated as about 300 K from the temperature dependence of the residual magnetization $M_r$ which is shown in Fig. 6a. As it is seen from Fig. 6b for sample 4, the Curie temperature is also slightly higher than 300 K. The experimental data presented in these figures are fitted with the theoretical expression obtained in Sec. 5. For the $Si_{1-x}Mn_x$/GaAs sample 7, of larger thickness, magneto-optical Kerr effect measurements were performed, showing hysteresis with $H_c$ about 0.7 kOe at $T \leq 80$ K, (see Fig.7).

The anomalous contribution to the Hall resistivity $\rho_{xy}^a(B)$ could be obtained from $\rho_{xy}(B)$ by subtraction of the normal component of the Hall effect which in turn could be found at magnetic fields above the saturation. From the magnetization values we determined the coefficient of the AHE, $R_s = \rho_{xy}^a/M$, which is about $1 \cdot 10^{-8}$ Ω·cm/G for sample 2 ($Al_2O_3$ substrate) at $T = 80$ K. A similar value $R_s \sim 0.7 \cdot 10^{-8}$ $\Omega \cdot$cm/G was obtained for sample 4 (GaAs substrate) at room temperature. The AHE angle tangent $\beta = \rho_{xy}^a / \rho_{xx}$ at 200 K is about $5 \cdot 10^{-3}$ being the hint of the strong spin polarization of carriers. If this were not the case, to observe a value $\beta \approx 5 \cdot 10^{-3}$ in the AHE angle tangent, taking into account the low carrier mobility (5 cm$^2$/V·s), an internal magnetic field $\approx 10$ T would be needed, which is unrealistic for the system with magnetic inclusions [3].

## 5. Discussion of the experimental results and theoretical model.

The experimental results obtained in Secs. 2-4 clearly demonstrate that FM order with an effective magnetic moment per Mn atom ($\approx 0.2\mu_B$/Mn) was observed at fairly high temperatures in Si$_{1-x}$Mn$_x$ alloys with a large manganese content ($x \approx 0.35$). The origin of this FM order is, however, not evident and has to be discussed.

The AHE (see Figs. 2 and 3) and magnetization (Figs. 4 and 5) properties, as well as the high values of $T_c$, could not be attributed to the bulk manganese silicides Mn$_n$Si$_m$ with ratio ($m/n$) ranging between 1.70 and 1.75 (for example, Mn$_4$Si$_7$, Mn$_{11}$Si$_{19}$, Mn$_{26}$Si$_{45}$, Mn$_{15}$Si$_{26}$, Mn$_{27}$Si$_{47}$). Indeed, the Curie temperature $T_c$ in such materials does not exceed 50 K and the effective magnetic moment per Mn atom is extremely small (for example $\approx 0.012\mu_B$/Mn in Mn$_4$Si$_7$ [8], i.e., significantly lesser than in the alloys under study). In these silicides the 3$d$-states of Mn are strongly hybridized with the 4($s,p$)-states of Si, so the spin density on the Mn atom is almost completely delocalized. Therefore, the materials are specified as exchange-enhanced paramagnets or weak itinerant ferromagnets. First principle calculations showed that the different phases Mn$_n$Si$_m$ are semiconducting, metallic, or half-metallic [9]. For example, spin-polarized calculations for Mn$_{11}$Si$_{19}$, Mn$_{15}$Si$_{26}$, and Mn$_{27}$Si$_{47}$ revealed that these phases are half-metallic, with full spin polarization of holes at the Fermi level. On the contrary, the ideally stoichiometric and unstressed Mn$_4$Si$_7$ is shown to be a semiconductor, with indirect band gap, although small non-stoichiometry or stress can lead to the closure of the gap, transforming the material into a metal. Due to the helicoidal long-period character of FM order in Mn$_n$Si$_m$ phases (see Refs.[14, 21]), a hysteresis loop in them should be absent or very smooth even at $T < T_c$. Note also, that the temperature dependence of the resistivity $R_{xx}(T)$ in Mn$_4$Si$_7$ differs drastically from the one we have observed. In fact, for Mn$_4$Si$_7$ the value $R_{xx}$ diminishes by more than 50 times in the temperature range between 300 K and 80 K and saturates at $T \leq 20$ K. In this case the ratio $R_{xx}(290K)/R_{xx}(5K)$ reaches 360, while in our sample the maximum of this ratio is equal only $\approx 2$. Furthermore, in our samples, $R_{xx}$ weakly depends on temperature in the interval 80K < T < 300K, and $R_{xx}(T)$ abruptly falls down at $T \leq$ 40 K for samples prepared on $Al_2O_3$ substrate (see Fig. 1).

It is quite clear that our $Si_{1-x}Mn_x$ films have strong structural disorder; particularly, their crystal lattice is far from displaying the regular periodicity of the bulk silicide $Mn_nSi_m$. The lack of local structural order around the Mn site can provide partial localization of Mn $3d$-states, thereby, the $Si_{1-x}Mn_x$ material is believed to contain Ångstrom sized magnetic defects (single Mn ions or molecular complexes containing Mn), denoted henceforth by the symbol $Mn_D$. We suppose, for concreteness, that these defects have magnetic configurations similar to that of $Mn_T$ centers, ($Mn_{Si}$-$Mn_T$) or ($Mn_T$-$Mn_T$) dimers, which are formed by Mn atoms being, correspondingly, in the substitutional ($Mn_{Si}$) and tetrahedral interstitial ($Mn_T$) positions in the Si lattice [17], where they have an effective magnetic moment $\sim (2-3)\mu_B$ per Mn atom.

So, there are serious reasons to distinguish two different components in the spin density of $Si_{1-x}Mn_x$ alloys: an itinerant (delocalized) component inherent to a weak $Mn_nSi_m$ ferromagnet and a localized component specific of the $Mn_D$ defects. According to the suggestion that our system consists of the $Mn_nSi_m$ matrix and of the $Mn_D$ defects inside it, we may assume the chemical formula of our material as $(Mn_nSi_m)_{1-\lambda}(Mn_D)_\lambda$. For concreteness, let us consider the $Mn_nSi_m$ host as the $Mn_4Si_7$ silicide. Thus, the $Si_{1-x}Mn_x$ alloy with nominal Mn content $x \approx 0.35$ could be formally regarded as a material with the formula $MnSi_{1.86}$, while $Mn_4Si_7$ could be represented as the $MnSi_{1.75}$ silicide. The identification $MnSi_{1.86} = (Mn\ Si_{1.75})_{1-\lambda}(Mn_D)_\lambda$ can easily explain the difference between an effective magnetic moment per Mn atom observed in our films ($\approx 0.2\ \mu_B$/Mn) and that observed in $Mn_4Si_7$ ($\approx 0.012\mu_B$/Mn). The effective magnetic moment per Mn atom in the $M_D$ defect is about $2.54\mu_B$/Mn for the $Mn_T$ center, $2.0\mu_B$/Mn for the ($Mn_{Si}$-$Mn_T$) dimer and $2.7\mu_B$/Mn for the ($Mn_T$-$Mn_T$) dimer, correspondingly [17], being somewhat less than the "nominal" value $\sim(4-5)\mu_B$/Mn for Mn atoms in the GaAs host [2]. Having this value and the measured effective magnetic moment per Mn atom, the amount of Mn atoms which do not belong to the host matrix and instead form magnetic defects could be evaluated as $\approx$ 8-10% of the total content of Mn in the $Si_{1-x}Mn_x$ film. For the $Si_{1-x}Mn_x$ alloy with $x \approx 0.35$ and total Mn concentration, $N_{Mn} \approx 2 \times 10^{22} cm^{-3}$, the concentration of magnetic defects, $N_{Mn}^D \approx (0.8-1.8) \times 10^{21} cm^{-3}$, could be estimated. This value corresponds to the mean distance between defects, $a_0 \approx (10-12)$ Å. One also could estimate the number of Si atoms per molecular complex $Z_{Si}^D \approx$ 4-5. Obviously, our estimates are rough and the different properties of samples prepared on various substrates may be accounted for by the crystal structure of the matrix imposed by the substrate. In turn, variations in the matrix structure could affect the concentration, size and shape of magnetic defects.

Obviously, there exist serious problems to explain high temperature ferromagnetism in our system. Indeed, so small a concentration of defects carrying magnetic moments is inadequate to promote FM order at high temperatures in the frame of the RKKY/Zener model. The main reason is that the above estimated mean distance between magnetic defects, $a_0 \approx (10-12)$ Å, is on the order of the period of RKKY oscillations: $2k_F a_0 \geq 5$, where $k_F^{-1} \leq 4$ Å is the inverse Fermi wave-vector. So, the spin glass regime would be more realistic in this situation, at odds with the observed FM behavior.

Below we discuss a possible model explaining FM order at high temperatures in our system. We presume that defects with local magnetic moments are embedded in the host, which is a weak itinerant ferromagnet, where strong spin fluctuations ("paramagnons") exist far above the "intrinsic" Curie temperature. We suggest that the Stoner enhancement of the exchange interaction between magnetic defects takes place, induced by spin fluctuations in the host; as a result, significant increase of the "global" Curie temperature of the system appears. The general possibility of such an enhancement in GaMnAs DMSs was firstly mentioned in Ref. [29]. Here, we use a simple phenomenological approach to describe this effect in metallic $Si_{1-x}Mn_x$ alloys.

For the sake of completeness, we recall some results of the theory of spin-fluctuation mediated itinerant ferromagnetism, which are relevant for the forthcoming analysis. In the mean-field approximation, the critical temperature of FM transition in the $Mn_nSi_m$ host (presumed metallic), $T_c^h$, is equal to $T_c^{MF}$ and can be found from the Stoner condition, $1 - U\chi^0(T_c^{MF}) = 0$. Here $U$ is effective potential of an exchange interaction between electrons, $\chi^0(T)$ is the response function of non-interacting electrons at the temperature $T$. For qualitative estimations of $T_c^{MF}$, one use a simple one-band model assuming a mean electron density of states $\bar{\rho} \approx W^{-1}$ in the interval of electron energy $0 < \varepsilon < W$, where $W$ is the electron band width. Presuming that the inequality $T << W$ ($W \approx 1.0 - 2.0$ eV, $T \leq 500$ K) is obeyed in the temperature range under study, one use the expansion $\chi^0(T) \approx \chi^0(0) - \Theta T^2$, were $\Theta \approx \bar{\rho} W^{-2}$ and obtain for the temperature $T_c^h$:

$$T_c^h = T_c^{MF} \approx \sqrt{\frac{|\alpha_{MF}|}{\bar{\rho}}} W, \quad \alpha_{MF} = U^{-1} - \chi^0(0) < 0. \qquad (1)$$

Thus, at $|\alpha_{MF}| \ll \overline{\rho}$ the Curie temperature $T_c^h$ is small with respect to the band width $W$ even in the mean-field approach. However, this approach is very rough and largely overestimates the Curie temperature, often predicting ferromagnetism in cases where no ferromagnetism is possible. So, the quantity $T_c^{MF}$ should be considered only as a certain characteristic temperature scale. Following Moriya`s approach [30], thermodynamical fluctuations of the spin density play a crucial role in itinerant ferromagnetism, significantly decreasing the actual transition temperature $T_c^h$ with respect to the mean-field value $T_c^{MF}$. Within the standard Murata-Doniach method [31], one adopts as a starting point the Landau expansion of the free-energy functional $F_h[\mathbf{m}]$ of the system, where $\mathbf{m}(\mathbf{r})$ is the classical vector characterizing the magnetization of itinerant electrons in the host,

$$F_h[\mathbf{m}] = \int d\mathbf{r} [\alpha_{MF}\mathbf{m}^2(\mathbf{r}) + \beta\mathbf{m}^4(\mathbf{r}) + \gamma\left(\frac{\partial \mathbf{m}}{\partial \mathbf{r}}\right)^2]. \qquad (2)$$

The coefficients $\alpha_{MF}$, $\beta$ and $\gamma$ in the Eq. (2) are almost independent on the temperature: $\alpha_{MF} = -|\alpha_{MF}|$ is a negative quantity defined in Eq. (1), while $\beta$ and $\gamma$ are positive ($\beta \approx \overline{\rho}W^{-2}$, $\gamma \approx \overline{\rho}\xi_0^2$, $\xi_0 = v_F/W$, $v_F$ being the electron velocity at the Fermi surface). We assume that $|\alpha_{MF}| \ll \overline{\rho}$; thus, in the mean-field approximation the functional (2) has the minimum at $m = m_0 = \sqrt{|\alpha_{MF}|/2\beta} \ll W$. To qualitatively describe the thermodynamics of the functional (2), one considers the fluctuations of magnetization $\mathbf{m}(\mathbf{r})$ in the Gaussian approximation. Separating the mean-field $\varphi(\mathbf{r})$ and spin fluctuation $\eta(\mathbf{r})$ components of the order parameter, $\mathbf{m}(\mathbf{r}) = \varphi(\mathbf{r}) + \eta(\mathbf{r})$, and averaging the functional over the random vector variable $\eta(\mathbf{r})$, one redefines the effective free energy functional of the host as

$$F_h[\varphi] = \int d\mathbf{r}\left[\alpha\varphi^2(\mathbf{r}) + \beta\varphi^4(\mathbf{r}) + \gamma\left(\frac{\partial\varphi}{\partial\mathbf{r}}\right)^2\right], \quad \alpha = \alpha_{MF} + \alpha_{SF}, \qquad (3)$$

$$\alpha_{SF} = \frac{5\beta T Q a^3}{\gamma\pi^2}\left[1 - \frac{arctg(\zeta Q)}{\zeta Q}\right]. \qquad (4)$$

The wave vector $Q$ determines an effective number of fluctuation modes and is, strictly speaking, a temperature-dependent quantity, $a$ denotes the lattice constant of the host. The phase transition temperature, $T_c^h$ is now defined by a condition $\alpha = 0$. From this condition one obtains:

$$T_c^h = T_c^{SF} \approx \frac{|\alpha_{MF}|T_0}{\gamma Q^2}, \qquad (5)$$

$T_0 = \frac{\pi^2 \gamma^2 Q}{5\beta a^3} \propto v_F Q \leq W$. Simple evaluation shows that $T_c^{SF} \propto (T_c^{MF})^2 / v_F Q \ll (T_0, T_c^{MF})$.

Note, that at $T > T_c^h$, the correlation length $\zeta(T)$ appears to be essentially renormalized with respect to the value obtained in the mean-field approximation $\zeta_{MF} = \zeta_0(\sqrt{W|\alpha_{MF}|})^{-1}$. One can see that, the regime $\zeta Q \gg 1$ is achieved in the temperature region of interest, $T_c^h < T \ll T_0$, where

$$\zeta Q \approx \sqrt{\frac{T_0}{(T - T_c^h)}} \qquad (6)$$

The Murata-Doniach approach [31] has mainly a methodological character, since it largely overestimates the role of long-range spin fluctuations at low temperatures, $T \ll T_c^h$, and introduces a systematic error near the Curie temperature, $T_c^h$. This problem is well known and rather severe but, on the other hand, the approach is qualitatively acceptable in the temperature region $T \gg T_c^h$, which is the aim of our analysis (see detailed discussion in Ref.[30]).

The presence of magnetic defects inside an itinerant FM host can significantly enhance the estimate (5) for the critical temperature of FM order in the system. At temperatures $T > T_c^h$, we shall treat these defects as point defects with local moments, which become centers for the formation of local regions with short-range FM order inside the host. Below, to describe this type of magnetic order we use the well-known concept of "local phase transition" [32] with a macroscopic, but finite, correlation length of spin fluctuations.

We consider the functional $F_h[\varphi]$ in Eqs. (3) above the phase transition temperature of the host $T_c^h$ (i.e. at $\alpha > 0$) as an effective mean-field type functional with a redefined order parameter $\varphi(\mathbf{r})$. Following the "local phase transition" concept, let us introduce the term $F_D[\varphi] = \int d\mathbf{r}\, \lambda(\mathbf{r})\varphi(\mathbf{r})$ representing a perturbation caused by the magnetic point defects dissolved in the host, where $\lambda(\mathbf{r})$ is the effective exchange potential. Minimizing the free energy functional of the system, $F[\varphi] = F_h[\varphi] + F_D[\varphi]$ with respect to the order parameter $\varphi(\mathbf{r})$, one obtains the self-consistency equation:

$$\gamma \frac{\partial^2 \varphi(\mathbf{r})}{\partial \mathbf{r}^2} - \tilde{\alpha}\varphi(\mathbf{r}) - 2\beta\varphi^3(\mathbf{r}) = \frac{1}{2}\lambda(\mathbf{r}). \qquad (7)$$

In the case of a single magnetic defect, placed at the point $\mathbf{r} = 0$, presuming that effective radius of the potential $\lambda(\mathbf{r})$ is small with respect to the correlation length $\zeta$, we can use the "point defect" approximation for $\lambda(\mathbf{r})$ and write $\lambda(\mathbf{r}) = \kappa \mathbf{S} \delta(\mathbf{r})$, where $\kappa$ is an exchange coupling integral ($\kappa \propto J_{pd} \overline{\rho}$ in the model of an indirect exchange between itinerant electrons and local moments, where $J_{pd}$ is the matrix element of this exchange), $\mathbf{S}$ is the vector of magnetic moment of the defect. If we omit the term $\sim \varphi^3(\mathbf{r})$ in Eq. (7), which is correct at $r \gg \zeta$, then the solution of Eq. (7) can be written as

$$\varphi_0(\mathbf{r}) = \frac{\kappa \mathbf{S}}{2\gamma} G(\mathbf{r}, 0). \tag{8}$$

$$G(\mathbf{r}, \mathbf{r}') = -\frac{1}{4\pi} \frac{\exp(-|\mathbf{r} - \mathbf{r}'|/\zeta)}{|\mathbf{r} - \mathbf{r}'|},$$

$G(\mathbf{r}, \mathbf{r}')$ is the Green function of the differential operator $\frac{\partial^2}{\partial \mathbf{r}^2} - \zeta^{-2}$. Eq. (8) describes the spin density redistribution in the host around a single magnetic defect. The characteristic radius of this redistribution coincides with the renormalized correlation length $\zeta(T)$ of the spin fluctuations in the host.

Let us now introduce in the host the set of magnetic point defects

$$\lambda(\mathbf{r}) = \sum_i \kappa \mathbf{S}_i \delta(\mathbf{r} - \mathbf{R}_i),$$

$\mathbf{R}_i$ and $\mathbf{S}_i$ are the random position and magnetic moment for $i$th defect, respectively. Omitting technical details (see Ref. [33]), one can show that, in the frame of a "point defect" approximation and to the second order in the exchange coupling integral $\kappa$, the contribution to the free energy of the effective coupling between magnetic defects in the host has the form:

$$F_{ex}^{SF} = \frac{1}{2} \sum_{i \neq j} J_{ij}^{SF} \mathbf{S}_i \mathbf{S}_j, \tag{9}$$

$$J_{ij}^{SF} = J^{SF}(|\mathbf{R}_i - \mathbf{R}_j|) = -\frac{\kappa^2}{4\pi\gamma} \frac{\exp(-a_0/\zeta)}{a_0},$$

where $a_0 = |\mathbf{R}_i - \mathbf{R}_j|$ is the distance between defects. The coupling integral $J_{ij}^{SF}$ is always ferromagnetic and has an exponential falloff ($|J^{SF}| \propto \frac{\exp(-a_0/\zeta)}{k_F a_0}$) at large distances, $a_0 \geq \zeta$. Thus, at extremely low concentration of defects their coupling through spin fluctuations seems to be negligible. However, at intermediate distances, $a_0 \propto \zeta \gg k_F^{-1}$ the contribution (9) may exceed or be comparable in order of value to the integral $J^{RKKY}$ in the RKKY mechanism of exchange coupling ($\left|J^{RKKY}\right| \propto (k_F a_0)^{-3}$). This means that, even in systems with a relatively low concentration of defects and obviously with a higher concentration ($a_0 \leq \zeta$), the contribution (9) has to be taken into account.

In order to evaluate the temperature of the FM ordering of local magnetic moments, $T_c^g$ (we call it the *global* Curie temperature), one can use the Weiss molecular field approximation:

$$T_c^g = \frac{S^2}{3k_B} J_0^{SF}(T_c^g), \qquad (10)$$

$$J_0^{SF}(T_c^g) = -\sum_j J_{ij}^{SF}(T_c^g) \approx N_{Mn}^D k^2 \zeta^2 / \gamma.$$

Hence, using Eq. (6) at $T_c^g >> T_c^h$ we obtain the estimate:

$$T_c^g \approx \sqrt{k^2 S^2 N_{Mn}^D T_0 \Big/ 3\gamma Q^2 k_B} \ . \qquad (11)$$

In our experiments $T_c^g \approx (300-400)K$, $T_c^h \approx 50K$, i.e., the temperature $T_c^g$ is significantly larger than $T_c^h$ and the correlation length $\zeta(T_c^g)$ may be evaluated as $\zeta(T_c^g) \approx Q^{-1}\sqrt{T_0/T_c^g} > k_F^{-1}\sqrt{W/T_c^g} \propto (3-4)k_F^{-1}$. The mean distance between magnetic defects is $a_0 \approx (10-12)\text{Å}$, so if we take $k_F^{-1} \approx (3-4)\text{Å}$, the regime $a_0 \propto \zeta >> k_F^{-1}$ is easily achieved. It can be easily shown that in this regime: $J_0^{SF} \propto (J_{pd}\overline{\rho})^2 / |\alpha|$, $J_0^{RKKY} \propto N_{Mn}^D J_{pd}^2 \overline{\rho}$, $\left|J_0^{SF}(T_c^g)/J_0^{RKKY}\right| \propto \overline{\rho}/|\overline{\alpha}| \propto W^2/[T_0(T_c^g - T_c^h)] >> 1$. Thus, the temperature $T_c^g >> T_c^h$ may be estimated in our model as $T_c^g \propto W\sqrt{T^{RKKY}/T_0} >> T^{RKKY}$, where $T^{RKKY} \propto N_{Mn}^D J_{pd}^2 \overline{\rho} \frac{S^2}{3k_B}$. Even if we take $T^{RKKY} \approx 10$ K at very low concentration of defects $N_{Mn}^D$, for $W \approx 10^4$ K, $T_0 \approx (10^3 - 10^4)$K, we obtain $T_c^g \approx (300-500)$K.

6. **Comparison of theoretical predictions with experimental results.**

First of all, let us point out that model of exchange (henceforth called SF model) presented above yields an estimated Curie temperature which is in agreement with experimental values, contrary to the standard RKKY/Zener model of exchange which predicts the spin glass regime. The SF model leads to a sizable growth of the Curie temperature in our system, with respect to the case when only the standard RKKY-like mechanism is taken into account.

In the frame of the SF model, the temperature dependence of the magnetization $M(T)$ should differ from that obtained within the RKKY theory. In the RKKY model, the mean-field value of exchange integral $J^{RKKY}$ does not depend on temperature, while the SF model yields $J^{SF}(T) \propto (T - T_c^h)^{-1}$. The equations describing the temperature dependence of the mean magnetization, $M(T)$, contain the factors $J^{RKKY}/kT$ in the frame of RKKY model or $J^{SF}(T)/kT$ in the frame of SF model, correspondingly. So, if the RKKY model fits the $M(T)$ dependence by the function $F(T/T_c)$, then the SF model exploits the same function, but with a different argument, $T(T - T_c^h)/T_c^g(T_c^g - T_c^h)$, to fit the $M(T)$ curve; as a result, for $T > T_c^h$ we have: $M(T)/M(T_c^h) \approx F[T(T - T_c^h)/T_c^g(T_c^g - T_c^h)]$.

It is known that, in DMSs the spatial disorder modifies the $M(T)$ dependence [34] (in particular, for the case of standard RKKY theory see Ref. [35]). Under the effect of disorder, the $M(T)$ dependence differs from that described by the Brillouin equation and could be fitted by the function $F(y)=1-y^n$ with $y = (T/T_c)^n$ (in particular, $n \approx 2$ for GaMnAs [36]). In the SF model, the experimental dependence $M(T)$ can be fitted with the same function, $F(y)$, but with $y = T(T - T_c^h)/T_c^g(T_c^g - T_c^h)$ (see Figs. 6a and 6b). Taking $n =1.3$-$1.5$ and $T_c^h = 50K$, we obtain the *fitted* Curie temperature $T_c^g(fitted) \approx 330$ K for both samples 2, 4 in a good agreement with the prediction of the SF theory, $T_c^g \approx (300 - 500)K$.

It should be mentioned that it is very hard to explain the observed results of the AHE and magnetic measurements based on the bulk $Mn_nSi_m$ properties. On the contrary, these results, as well as the data of resistivity measurements, are in qualitative agreement with the proposed model of the sample structure: magnetic defects (FM molecular clusters) embedded in the $Mn_nSi_m$ matrix. Indeed, in the whole temperature range $R_{xx}(T)$ differs drastically from its behavior in $Mn_nSi_m$, furthermore at the temperature of magnetic ordering for the host Mn silicide, $T_c^h$, the $R_{xx}(T)$ dependence abruptly falls down. Also the sample characteristics depend on the substrate type which affects the matrix structure, but the main magnetic properties (magnetic moment, existence of the AHE) do not change significantly.

## 7. Conclusions

Room temperature ferromagnetism has been achieved in Si based structures with high Mn content. The important point is that the FM behavior was detected not only by magnetization measurements but also by the observation of the anomalous Hall effect. So, FM order involves charge carriers which are most probably at least partly spin polarized, and is not due to separated magnetic inclusions not interacting with carriers. The magnetic hysteresis loop as well as the temperature dependence of the saturation magnetization and coercive force measured by magnetic and transport methods are similar, and this fact proves the previous statement. The Curie temperature obtained from the temperature dependence of residual magnetization was found to be about 330 K. It is hard to explain the whole set of experimental results in the frame of the standard RKKY/Zener model of exchange between local moments of manganese, or by the formation of a weak itinerant FM (manganese silicide) in the $Si_{1-x}Mn_x$ ( $x \approx 0.35$ ) alloy under consideration. To explain the obtained experimental data, we used a more complex model of FM order, based on the conception of a two-phase magnetic material composed of defects with local magnetic moments, which are embedded in the host, assumed to be a weak itinerant ferromagnet. We argued that molecular clusters (probably, $Mn_{Si}$-$Mn_T$ or $Mn_T$-$Mn_T$ dimers), containing a minority of Mn atoms, form these defects in our alloy, while the majority of Mn atoms is involved in the formation of the $Mn_nSi_m$ host. The observed FM ordering at high temperatures (>300 K) is due to the Stoner enhancement of the exchange coupling between local moments of defects provided by strong spin fluctuations ('paramagnons') in the host. Our theoretical predictions and experimental results are in good qualitative agreement.

**Acknowledgements**

We thank Profs. C. Back, A.V. Vedyaev, A.B. Granovskii and E.Z. Meilikhov for fruitful discussions. V.V.T. thanks the Basque Foundation for Science (Ikerbasque) and Donostia International Physics Center (DIPC) for organizing and financial help. B.A. thanks Profs. S. Ganichev and C. Back for hospitality and the University of Regensburg for financial support. The work is partially supported by RFBR (grants 08-02-01462, 09-02-12108, 09-07-12151, 09-07-13594, 10-07-00492 and 10-02-00118).

# Figure captions

Fig. 1. The temperature dependence of resistivity for $Si_{1-x}Mn_x$ ($x \approx 0.35$) samples deposited on $Al_2O_3$ (a) and GaAs (b) substrates. The number at curves corresponds to the sample number. The growth temperature: 1, 2, 4, 5, 7 – 300 °C; 3 – 350 °C; 6 – 200 °C.

Fig. 2. The Hall effect resistivity hysteresis curves for the sample 2 ($Si_{1-x}Mn_x$ /$Al_2O_3$) at various temperatures; low temperatures (a), higher temperatures (b). Thick arrows show the magnetic field sweep direction. The inset presents the Hall effect resistivity curve at room temperature.

Fig. 3 The Hall effect resistivity versus magnetic field for the sample 4 ($Si_{1-x}Mn_x$ /GaAs) at various temperatures. In inset the room temperature data are shown.

Fig. 4. The magnetization hysteresis curves for the sample 2 ($Si_{1-x}Mn_x$/$Al_2O_3$) (a) and the sample 4 ($Si_{1-x}Mn_x$ /GaAs) (b) at various temperatures. Arrows show the magnetic field sweep direction. The hysteresis loops for the samples on GaAs substrate are considerably narrow than for $Si_{1-x}Mn_x$ /$Al_2O_3$ structures, therefore the measured $M(H)$ curves for the sample 4 have been fitted with the Langevin function for obtaining the coercivity $H_c$. Curves 1-3 correspond to the fitted dependencies $M(H)$ at 80, 200 and 300 K. The measured dependence $M(H)$ for 80 K is shown (curve 1').

Fig. 5. The normalized coercivity $H_c(T)/H_c(0)$ (a) and saturation magnetization $M_s(T)/M_s(0)$ (b) obtained from both transport (triangles) and magnetic measurements (circles) for the sample 2 ($Si_{1-x}Mn_x$ /$Al_2O_3$). $H_c(0)$ and $M_s(0)$ are the low (helium) temperature values. $M_s(T)$, $H_c(T)$ and $M_{sH}(T)$, $H_{cH}(T)$ magnetization and coercivity values calculated from magnetic measurements and Hall effect measurements, respectively.

Fig. 6. Temperature dependencies of normalized remanent magnetization $M_r(T)/M_r(0)$ for the sample 2 ($Si_{1-x}Mn_x$ /$Al_2O_3$) (a) and normalized saturation magnetization $M_s(T)/M_s(0)$ for the sample 4 ($Si_{1-x}Mn_x$ /GaAs) (b). For the sample 4 the $M_s(T)/M_s(0)$ dependence is presented instead of $M_r(T)/M_r(0)$ because the hysteresis loop for $Si_{1-x}Mn_x$ /GaAs samples is very narrow and the remanent magnetization could not be evaluated with high enough accuracy. The $M_s(T)$ value was measured at $B$= 0.5 T. Magnetizations $M_r(0)$ and $M_s(0)$ correspond to $T = 4.2K$. The solid lines are the fitting of temperature dependencies for $M_r(T)/M_r(0)$ and $M_s(T)/M_s(0)$ by theoretically obtained function $F(y)=1-y^n$ with $y = T(T - T_c^h)/T_c^g(T_c^g - T_c^h)$ and $T_c^h$ = 50K related to the presented model. Fitting parameters are $T_c^g$ = 330 K for both curves, and $n$ is 1.5 and 1.3 for sample 2 and sample 4 respectively.

Fig. 7. Hysteresis curves for normalized Magnetic Optical Kerr Effect (MOKE) for the sample 7 ($Si_{1-x}Mn_x$ /GaAs) taken at temperatures 80 K (curve 1, circles) and 200K (curve 2, triangles). Arrows show the magnetic field sweep direction. Inset shows the temperature dependence of the coercivity.

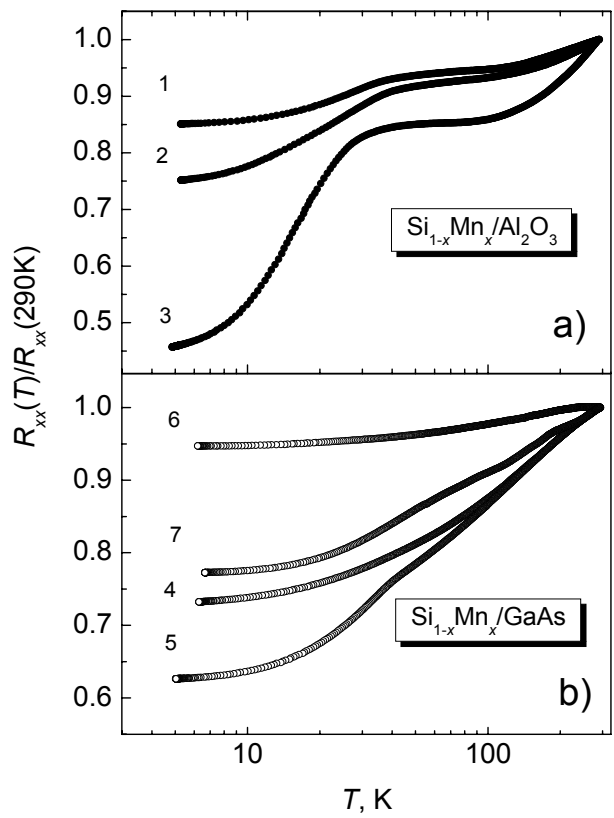

Fig.1.

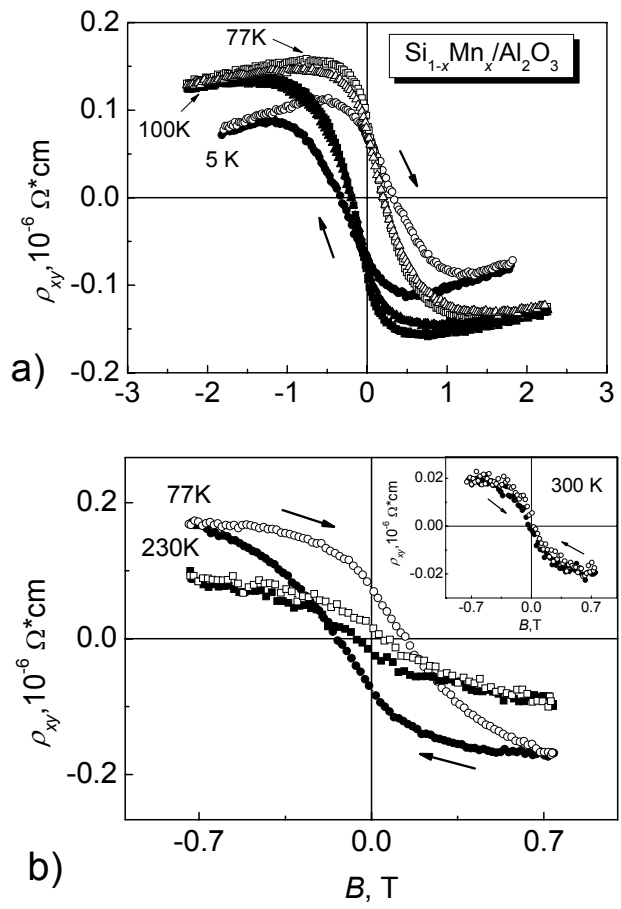

Fig.2

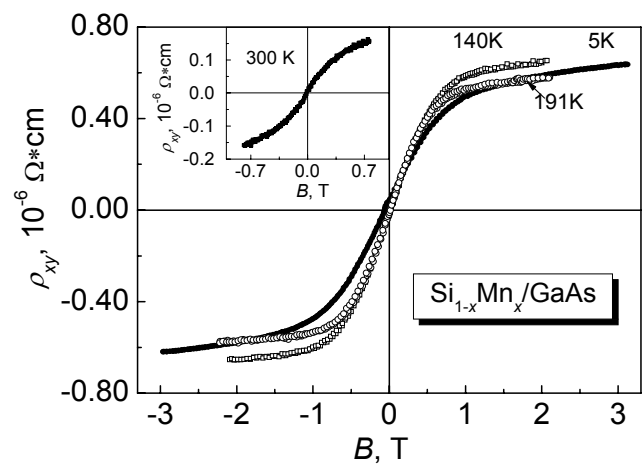

Fig. 3

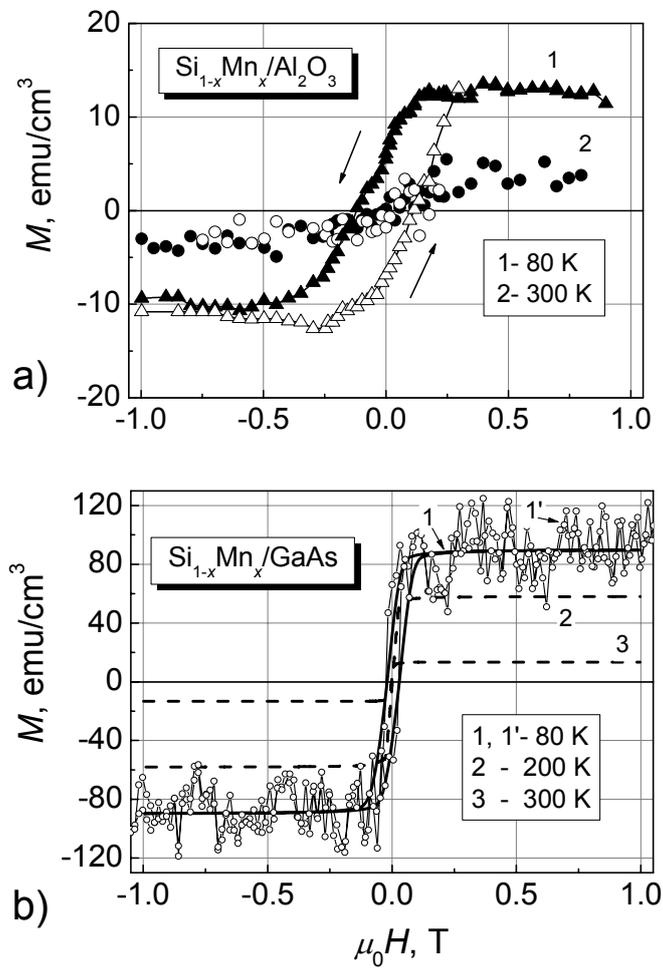

Fig.4

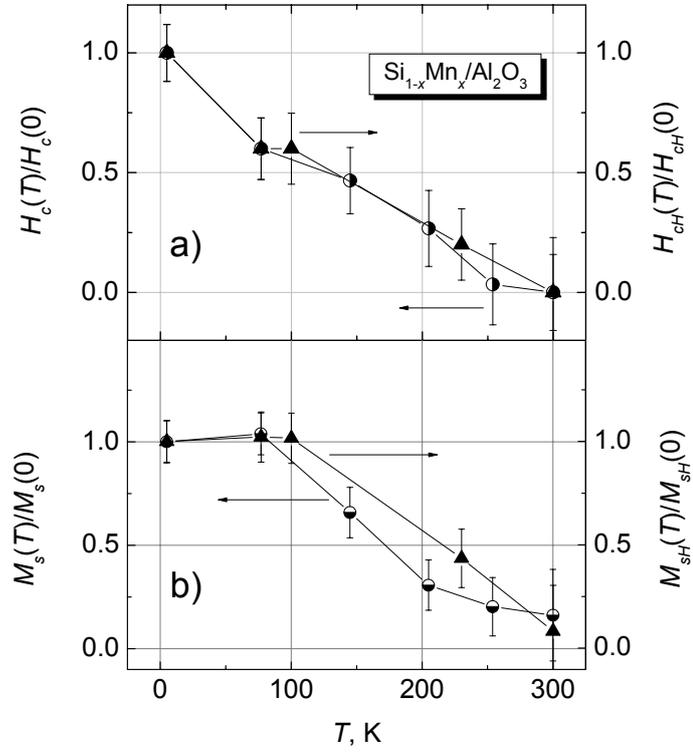

Fig.5.

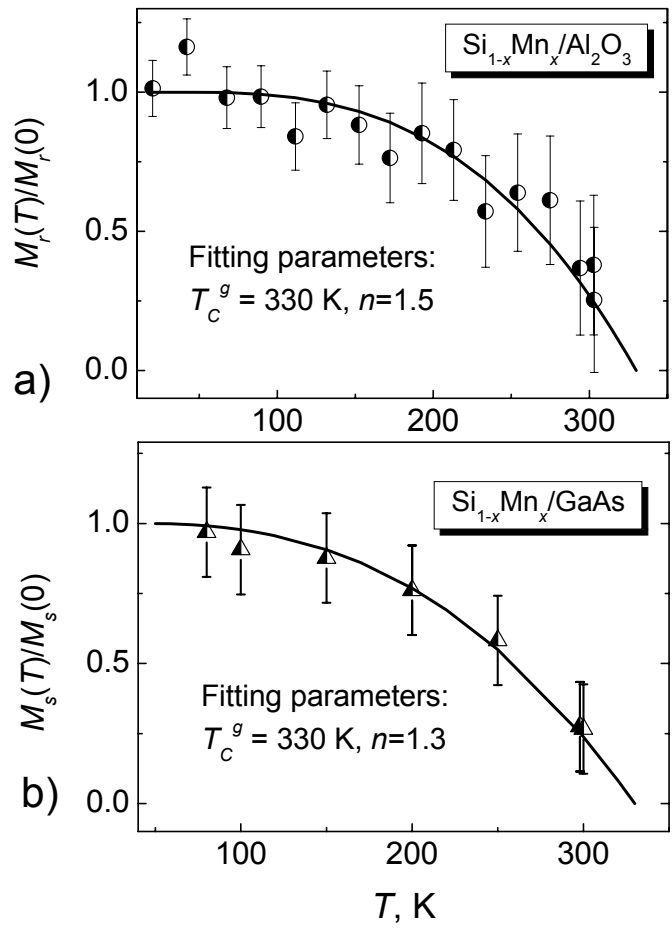

Fig.6

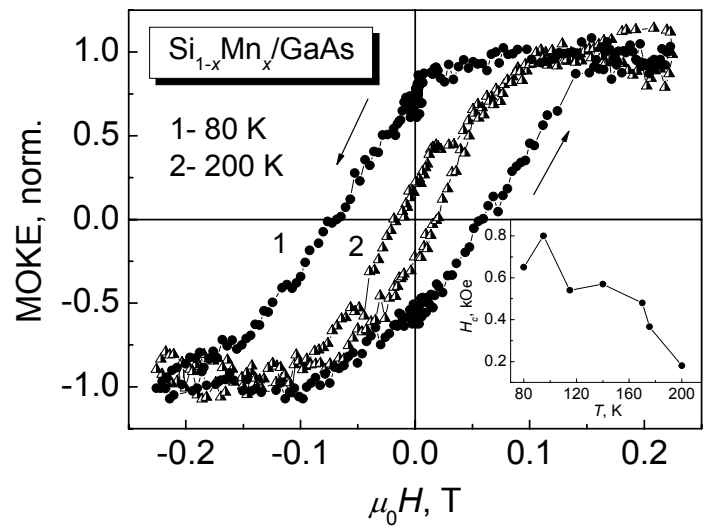

Fig.7